\documentclass[prl,twocolumn,showpacs]{revtex4}
\usepackage[dvips]{graphicx}
\usepackage{latexsym}
\usepackage{bm}
\begin{document}
\newcommand{\fig}[2]{\includegraphics[width=#1]{#2}}
\newcommand{\pprl}{Phys. Rev. Lett. \ } 
\newcommand{\pprb}{Phys. Rev. {B}} 
\newcommand{\be}{\begin{equation}}
\newcommand{\ee}{\end{equation}}
\newcommand{\bea}{\begin{eqnarray}} 
\newcommand{\eea}{\end{eqnarray}}
\newcommand{\nn}{\nonumber} 
\newcommand{\la}{\langle}
\newcommand{\ra}{\rangle} 
\newcommand{\dg}{\dagger}
\newcommand{\br}{{\bf{r}}} 
\newcommand{\tnh}{{\rm tanh}}
\newcommand{\sh}{{\rm sech}} 
\newcommand{\pll}{\parallel} 
\newcommand{\upa}{\uparrow} 
\newcommand{\dna}{\downarrow}
\newcommand{\lw}{\Lambda_\omega}
\newcommand{\lwp}{\Lambda_\omega^\prime}
\newcommand{\tphi}{{\tilde\phi}}
\newcommand{\ru}{{U}}
\newcommand{\rj}{{J}}
\newcommand{\rt}{{t}}
\newcommand{\rru}{{U_r}}
\newcommand{\rrj}{{J_r}}
\newcommand{\rrt}{{t_r}}

\title{Tunneling, dissipation, and superfluid transition in
quantum Hall bilayers}

\author{Ziqiang Wang}
\affiliation{Department of Physics, Boston College, Chestnut Hill, MA 02467}

\date{\today}

\begin{abstract}

We study bilayer quantum
Hall systems at total Landau level filling factor $\nu=1$ in the presence
of interlayer tunneling and coupling to a dissipative normal fluid.
Describing the dynamics of the interlayer phase by an effective
quantum dissipative $XY$ model, we show that there exists
a critical dissipation $\sigma_c$ set by the conductance
of the normal fluid. For $\sigma > \sigma_c$, interlayer
tunnel splitting drives the system to a $\nu=1$ 
quantum Hall state. For $\sigma <\sigma_c$, interlayer tunneling is
irrelevant at low temperatures, the system exhibits a superfluid transition to 
a collective quantum Hall state supported by spontaneous interlayer phase 
coherence. The resulting phase structure and the behavior of the
in-plane and tunneling currents are studied in connection to experiments.

\typeout{polish abstract}
\end{abstract}


\maketitle

A double-layer quantum Hall system at total Landau level filling factor
$\nu=1$ exhibits a novel weak to strong coupling phase transition as
a function of the interlayer separation $d/\ell$, where $\ell$ is the
magnetic length. The weak-coupling (WC) phase at large $d/\ell$ corresponds
to two weakly coupled compressible fluids
with unquantized Hall conductance, whereas the strong-coupling (SC)
phase at small $d/\ell$ is a $\nu=1$ quantum Hall liquid
\cite{eisenstein,murphy,book,girvin}. If the layer
index is regarded as a pseudospin index, the SC
phase exhibits pseudospin ferromagnetism. Experimental evidence
in support of spontaneous interlayer coherence in the SC
phase was discovered recently. The interlayer tunneling
conductance shows a sharp zero-bias peak reminiscent of the DC 
Josephson effect \cite{spielman}, consistent with
spontaneous breaking of the gauge symmetry 
associated with the {\it charge difference} between the 
layers \cite{fertig,wen}. The fact that the zero-bias 
peak maintains a finite height and a nonzero width at low 
temperatures signals the importance of dissipation
and generates a debate on whether there is true
superfluidity in the presence of interlayer tunneling \cite{joglekar}.
Considerable theoretical efforts have been devoted to understanding
the effects of dissipation on the tunneling conductance
\cite{balents,stern,fogler,fertig2} and the nature of the weak to strong
coupling transition \cite{joglekar,schliemann,kim,sheng}.

In this paper, we study the effects of interlayer tunneling on the
(pseudospin) superfluid transition in the presence of dissipation in realistic
samples. The source of dissipation we consider comes from the
spatial fluctuations in the electron density near the transition 
\cite{spielman}. As a consequence of such fluctuations, the width of
the longitudinal drag resistance peak across a Hall drag transition
is nonzero at low temperatures \cite{kellogg}.
We thus view the bilayer near the SC to WC transition as an
inhomogeneous mixture of SC droplets 
immersed in the background of the WC fluid. This picture was
introduced recently by Stern and Halperin \cite{sternhalperin} to describe the 
enhancement of the interlayer Coulomb drag near the
transition \cite{kellogg}. In that work, the transition and the 
associated transport properties were treated in terms of classical percolation 
of the SC droplets.
Here, we study the effects of quantum fluctuations by exploring the
analogy to granular superconductors near the
superconductor-insulator transition \cite{chakravarty,fisher,abeles,
doniach,efetov,simanek}.
Specifically, we consider a two-dimensional array of SC
droplets embedded in the dissipative environment of the
background diffusive normal fluid.
The SC droplets are coupled by the Josephson coupling $J$, and
to the WC dissipative normal fluid capacitively. After integrating out
the WC electrons, we arrive at a dissipative, quantum $XY$ model
for the dynamics of the phases on the SC droplets.
The interlayer tunneling $t$ plays the role of a symmetry breaking field
which, in the absence of dissipation, is relevant in the renormalization
group (RG) sense and destroys the putative superfluid transition.
In the presence of dissipation, characterized by the pseudospin conductance
$\sigma$ of the normal fluid, we find that quantum phase fluctuations
lead to a critical dissipation $\sigma_c$. For $\sigma >\sigma_c$, $t$ is
{\it relevant} and flows to large values at low
temperatures. The system develops a mass gap associated
with the symmetric-antisymmetric tunnel splitting of the
single-particle band and is in the $\nu=1$ integer quantum Hall state.
For $\sigma < \sigma_c$, the interlayer tunneling is
{\it irrelevant}, $t$ scales to zero in the low temperature limit.
We show that in this case the system 
undergoes a zero-temperature superfluid transition to
a collective quantum Hall state with spontaneous
interlayer phase coherence when $J$ exceeds
a critical value $J_c$. The tunneling current is found to
exhibit non-monotonic temperature dependence near the
transition.

We begin with the action for a Josephson junction array of SC droplets
(setting $\hbar=1$),
\bea
S_{sc}[\phi]\!\!&=& \!\!\int_0^{\beta}\!d\tau\biggl[{1\over2U}\sum_i
\left({\partial\phi_i\over\partial\tau}\right)^2+J\sum_{\langle i,j\rangle}
\left(1-\cos\Delta\phi_{ij}\right)
\nonumber \\
&-t&\sum_i\cos\phi_i(\tau)\biggr],
\label{ssc}
\eea
where $\phi_i$ is the phase of the interlayer excitonic order parameter
on the $i$-th droplet, $\langle\psi_1^\dagger(r_i) \psi_2(r_i)
\rangle\propto e^{i\phi_i}$ with $\psi_\alpha^\dagger$ creating
an electron in layer-$\alpha$ \cite{book,girvin}. 
The nearest neighbor droplets are coupled by Josephson coupling
$J$ in Eq.~(\ref{ssc}), where $\Delta\phi_{ij}=\phi_i-\phi_j$, and
$U=e^2/C_s$ is the charging energy set by the droplet capacitance
$C_s\sim \varepsilon\xi^2/d$ with $\xi$
the average droplet size.
The last term in Eq.~(\ref{ssc}) represents the interlayer
tunneling $t=N_0\Delta_{SAS}$ with $\Delta_{SAS}$ the 
symmetric-antisymmetric band splitting
and $N_0$ the average number of electrons per droplet.

The SC droplets and the WC background fluid are not at the same
potential.
Denoting the phase of the WC electrons by $\varphi$ \cite{esa}, 
there is a voltage drop 
$(\partial\phi_i/\partial\tau-\partial\varphi_i/\partial\tau)/e$
across the 2D Thomas-Fermi screening
length $\lambda_{TF}=1/4\pi e^2 (dn/d\mu)$
around the $i$-th droplet, where $dn/d\mu$ is
the compressibility of the normal fluid.
The action governing their capacitive coupling is thus
\be
S_{sc-wc}[\phi,\varphi]
=\int_0^\beta d\tau\sum_i{1\over2}{C_w\over e^2}\left({\partial\phi_i
\over\partial\tau}
-{\partial\varphi_i\over\partial\tau}\right)^2.
\label{sscwc}
\ee
where the effective capacitance $C_w\sim 1/\lambda_{TF}$.
The phase of the dissipative normal fluid is
damped and follows the diffusive (Ohmic) dynamics,
\be
S_{wc}[\varphi]={1\over2\beta}\sum_n\sum_{\langle i,j\rangle}\sigma
\vert\omega_n\vert\vert\Delta\varphi_{ij}(\omega_n)\vert^2,
\label{swc}
\ee
where $\omega_n=2\pi n/\beta$ is the Matsubara frequency and
$\sigma$ is the pseudospin conductivity 
in units of $e^2/\hbar$. The total action of the system is
\be
S=S_{sc}[\phi]+S_{sc-wc}[\phi,\varphi]+S_{wc}[\varphi].
\label{stotal}
\ee

To obtain an effective action for the phase of the SC droplets, we integrate
out the phase $\varphi$ of the WC electrons
in Eq.~(\ref{stotal}) using Eqs.~(\ref{swc}) and (\ref{sscwc}).
The result is
\bea
S_{\rm eff} &=&
{1\over2\beta}\sum_q\sum_n\Pi(q,\omega_n)\vert\phi(q,\omega_n)\vert^2
\nonumber \\
&+&\!\!\!\int_0^\beta\! d\tau\biggl[J\sum_{\langle i,j\rangle}(1-\cos
\Delta\phi_{ij})-t\sum_i\cos\phi_i\biggr],
\label{seff} 
\eea
where
\be
\Pi(q,\omega_n)={\omega_n^2\over U}+{\sigma\Delta(q)\over
\vert\omega_n\vert+D\Delta(q)}\omega_n^2.
\label{pi}
\ee
In Eq.~(\ref{pi}), $D=e^2\sigma/C_w$ is the diffusion
constant, and $\Delta(q)=4-2[\cos(q_xa)+\cos(q_ya)]$ with $a$
the average distance between the SC droplets,
set to unity hereafter. 
Note that since dissipation implies that charges can be transferred
continuously, the phase $\phi$ is noncompact
\cite{chakravarty,fisher}, and the action (\ref{seff}) 
is in fact a dissipative quantum Sine-Gordon
model in 2+1 dimensions.

The phase fluctuations described by Eq.~(\ref{pi})
have two very different dynamic regimes.
(i) For fluctuations in the quantum regime, $\vert\omega_n\vert>D\Delta(q)$, 
\be
\Pi(q,\omega_n)={\vert\omega_n\vert\over U}\left[
\vert\omega_n\vert+\sigma U\Delta(q)\right],\quad
\vert\omega_n\vert>D\Delta(q),
\label{piquantum}
\ee
the phase dynamics is dissipative.
(ii) For $\vert\omega_n\vert<D\Delta(q)$, the level spacing
is larger than the frequency and dissipation is inactive. The system
returns to the charging regime with
\be
\Pi(q,\omega_n)=(C_s+C_w){\omega_n^2\over e^2}\equiv
{\omega_n^2\over U_r},\quad \vert\omega_n\vert<D\Delta(q).
\label{piclassical}
\ee
In this case, the WC normal fluid merely screens the electrostatic
interaction of the exciton-pairs. The dynamics of the
phase is again capacitive, with a renormalized charging energy
$U_r$ determined by putting the capacitors in parallel.
As a result, {\it the phase space for the damping of the SC
phase is limited to the quantum fluctuation regime},
in contrast to resistive shunted superconducting grains
\cite{chakravarty,fisher}.
Writing
$
\phi(q,\omega_n)=h(q,\vert\omega_n\vert >Dq^2)
+\theta(q,\vert\omega_n\vert < Dq^2),
$ 
our strategy now is to integrate out the dissipative fluctuations $h$
in regime (i) to arrive at an effective theory of capacitive junctions
for $\theta$ in regime (ii) with renormalized parameters.
The correlation function 
$G_h({\bf r_i}-{\bf r_j},\tau-\tau^\prime)=\langle h_i(\tau)
h_j(\tau^\prime)\rangle$ is given by
\be
G_h({\bf r},\tau)={2\over\beta}\sum_{n\ge0}{U\over\omega_n}
\sum_{q<\Lambda_\omega}{\cos(\omega_n\tau)e^{i{\bf q}\cdot{\bf r}}
\over\omega_n+\sigma U\Delta(q)}.
\label{gh}
\ee
Here the cutoff $\Lambda_\omega$ for the momentum summation is specified by the
condition $D\Delta(\Lambda_\omega)=\vert\omega_n\vert$ 
in Eq.~(\ref{piquantum}).

Consider first the renormalization of the tunneling term in Eq.~(\ref{seff}).
Within the standard Gaussian approximation,
$
t\langle \cos\phi_i\rangle_h=t\langle\cos(h_i+\theta_i)\rangle_h
\equiv t_r\cos\theta_i.
$
The renormalized tunneling is given by
$t_r=t\exp(-W_t)$, where the ``Debye-Waller'' factor
\be
W_t={1\over2}\langle h_i^2(\tau)\rangle={1\over2}G_h(0,0),
\label{wt}
\ee
represents the renormalization
of the tunneling due to dissipative phase fluctuations.
Evaluating $G_h(0,0)$ using Eq.~(\ref{gh}) in the continuum limit shows
that $W_t$ is logarithmically divergent at low temperatures,
\be
W_t=-{1\over8\pi^2\sigma}\ln\left(1+C_w/C_s\right)\ln(T\tau_0),
\label{wtlog}
\ee
where $\tau_0$ is a microscopic, short time scale serving as
a high frequency cutoff. This singularity is a form of the {\it orthogonality
catastrophe}: the fast voltage fluctuations created by the coherent tunneling
of the electrons between the layers cannot be followed adiabatically 
by the dissipative dynamics. As a result, the renormalized
tunneling vanishes according to a power law,
\be
t_r=t(T\tau_0)^{2\sigma_c/\sigma},
\label{trpower}
\ee
at low temperatures, where $\sigma_c=(1/16\pi^2)\ln(1+C_w/C_s)$. 
The renormalization of $J$ can be obtained similarly by writing
$J\langle 1-\cos\Delta\phi_{ij}\rangle_h=J_r(1-\cos\Delta\theta_{ij})$,
with $J_r=J\exp(-W_J)$. However, the limited
phase space for the dissipative fluctuations renders
\be
W_J={1\over8\pi^2}\int_0^{1/\tau_0}{U\over\omega}\int_0^{\sqrt{\omega/D}}
qdq{q^2\over \omega+\sigma Uq^2},
\label{wj}
\ee
free of the infrared singularity and
$
W_J=(1/16\pi^2)[1-(C_s/C_w)\ln(1+C_w/C_s)](1/D\tau_0).
$
This is in sharp contrast to resistive
shunted Josephson junctions where the {\it unlimited} dissipative fluctuations
lead to logarithmic corrections to the Josephson coupling
\cite{chakravarty,fisher}. 

Eq.~(\ref{trpower}) suggests that the scaling dimension of
the tunneling $t$ is $-2z_T\sigma_c/\sigma$, where $z_T=2$ is the
thermal exponent relating the temperature scale to the length scale.
Taking into account its naive dimension, we obtain
the RG dimension for $t$, $D_t=2+z_T-
2z_T\sigma_c/\sigma=4(1-\sigma_c/\sigma)$. Thus,
$\sigma_c$ is identified with the critical dissipation
that separates a weak-dissipation phase with irrelevant interlayer tunneling
for $\sigma < \sigma_c$ from a strong-dissipation phase for 
$\sigma > \sigma_c$ where the interlayer tunneling is 
relevant. This conclusion can be made more explicit if the Debye-Waller
factor $W_t$ in Eq.~(\ref{wt}) is evaluated self-consistently by
including the mass term $\sqrt{Ut_r}$ in the $h$-field propagator  
in Eq.~(\ref{gh}), which replaces $T$ as the cutoff for the logarithmic
divergence in Eq.~(\ref{wtlog}) at low temperatures. We then obtain
$t_r=t(Ut_r\tau_0^2)^{\sigma_c/\sigma}$ at $T=0$
in place of Eq.~(\ref{trpower}).
This equation has a physical, nonzero solution,
$t_r=(1/\tau_0)(U\tau_0)^{\sigma_c/(\sigma-\sigma_c)}
(t\tau_0)^{\sigma/(\sigma-\sigma_c)}$ only for
$\sigma > \sigma_c$, whereas $t_r=0$ for $\sigma <\sigma_c$,
consistent with the conclusion drawn from the
RG dimension. The critical dissipation depends on
the ratio $C_w/C_s$ and is not universal.

The renormalized action for the phase $\theta_i$ in the charging regime 
has the same form as Eq.~(\ref{ssc}), but
with renormalized parameters $U_r$, $J_r$, and $t_r$.
We now study the development of interlayer phase coherence
characterized by a nonzero order parameter $\langle\cos\theta_i\rangle$.
It turns out that
the interlayer coherence can be driven by either the interlayer tunneling
or the Josephson coupling between the droplets,
depending on the degree of dissipation.
To illustrate this, we set $J=0$ which gives the Lagrangian
${\cal L}=(1/2U_r)(\partial\theta_i/\partial\tau)^2-t_r\cos\theta_i$.
The system thus turns into decoupled droplets, or single vertical junctions,
and the interlayer tunneling $t_r$ serves as the Josephson coupling.
The competition between the charging energy and the Josephson coupling
of a single junction in granular metals was studied by Abeles \cite{abeles}.
For the phase coherence to develop at $T=0$, the frequency of the
quantum zero point oscillations, $\omega_p=\sqrt{t_rU_r}$ (interlayer
Josephson plasma frequency) must not exceed the Josephson coupling 
strength $t_r$, i.e. $t_r > U_r$. 
This condition is satisfied for strong dissipation since the renormalized 
$t_r$ scales to large values at low temperatures. Thus for
$\sigma >\sigma_c$, the interlayer tunneling 
drives the layers phase coherent even in the absence of coupling
between the pseudospins in the 2D plane. 
The system is in an integer quantum Hall state due to the relevant
symmetric-antisymmetric tunneling splitting of the single-particle
band in this regime.

We now focus on the case of weak dissipation, $\sigma < \sigma_c$,
where the renormalized tunneling $t_r$ scales to zero
according to Eq.~(\ref{trpower}). In this case, the Abeles criterion
cannot be satisfied at low temperatures and interlayer tunneling
does not lead to phase coherence. However, we show that a zero
temperature pseudospin superfluid transition can occur through
Josephson coupling between the droplets.
Following Doniach \cite{doniach}, Efetov \cite{efetov}, and 
\^{S}im\'{a}nek \cite{simanek}, we replace $J_r\cos(\theta_i-\theta_j)
\to J_r\langle\cos\theta_i\rangle\cos\theta_j$ in Eq.~(\ref{ssc}). 
The array is thus described by an effective single-droplet 
theory specified by the self-consistency equation for the superfluid 
order parameter $\Psi=2\langle\cos\theta_i\rangle$,
\bea
\Psi&=&{2\over Z_0}\int{\cal D}[\theta]\cos\theta_j e^{-\int_0^\beta d\tau
{\sum_i}
L_0[\theta_i(\tau)]}
\label{psi} \\
L_0&=&
{1\over2 U_r}\left({\partial\theta_i\over\partial\tau}\right)^2
+2\Psi J_r\cos\theta_i-t_r\cos\theta_i,
\label{srmf}
\eea
where $Z_0$ is the meanfield partition function. 
Since the renormalized tunneling $t_r$ is small at low temperatures,
$\Psi$ can be obtained to leading order in 
$t_r=t(T\tau_0)^\mu$, $\mu=2\sigma_c/\sigma>2$ by
linearizing Eq.~(\ref{psi}),
\be
\Psi=t(T\tau_0)^\mu\chi(T),\quad \chi(T)={\chi_0(T)\over1-2J_r\chi_0(T)}.
\label{chi}
\ee
Here $\chi(T)$ is the RPA-like susceptibility and
\be
\chi_0(T)=2\int_0^\beta d\tau\langle\cos\theta(\tau)\cos\theta(0)\rangle.
\label{chi0}
\ee
Evaluating $\chi_0$ using the Gaussian fluctuation in Eq.~(\ref{srmf}),
$
\langle\phi(\tau)\phi(0)\rangle=(U_r/\beta)\sum_{n}
\cos(\omega_n\tau)/\omega_n^2,
$
We obtain 
\be
\chi_0(x)=\int_0^\beta d\tau e^{-{U_r\over2\beta}\tau(\beta-\tau)}
={8\over U_r}y(x), \ \  x=\sqrt{U_r\over8T},
\label{chi02}
\ee
where $y(x)=xe^{-x^2}\int_0^x dz e^{z^2}$,
a hypergeometric function, has the limiting behavior
$\lim_{x\to\infty}y(x)=1/2$. 
At $T=0$, the onset of spontaneous interlayer phase coherence is
determined by $2J_r\chi_0(\infty)=1$, leading to
a quantum critical point (QCP) at $J_r^c=U_r/8$, 
which separates a SC phase with spontaneous 
interlayer phase coherence for $J_r > J_r^c$ from an interlayer
incoherent WC phase for $J_r<J_r^c$. 
Since $U\sim (e^2/\varepsilon\ell)(\ell/\xi)^2(d/\ell)$ and 
$J\sim e^2/\varepsilon\ell$, 
this condition corresponds to the existence of a critical layer
separation $d_c/\ell$ below
which 2D superfluidity develops, in qualitative agreement 
with experiments \cite{murphy,spielman}. In the present
theory based on the droplet picture, the critical
layer separation is not universal, but depends on the 
average size of the droplets in the critical regime, a prediction
that can be verified by experiments. 

At finite temperatures, since the symmetry-breaking field $t_r$ is
finite, one does not expect phase transitions in the usual sense. However,
the divergence of the susceptibility $\chi$ at finite temperatures when
$1=2J_r\chi_0(T_c)$ would signal the onset of rapid growth of 
the magnetization $\Psi$ associated with a crossover or
a metamagnetic-like transition.
The effects of a finite but small $t_r$ on $T_c$
near the QCP can be estimated from the nonlinear susceptibility
by including the contributions from interlayer tunneling $t_r$ in
$\chi_0(T)$. To leading order in $t$ and at low temperatures, we obtain
\be
T_c\simeq{1\over\tau_0}\left[\left({3\over2 t\tau_0}\right)
\left({J_r^c-J_r\over J_r}\right)\right]^{1/(\mu-1)},\quad J_r < J_r^c.
\label{tcfinal}
\ee
Although the tunneling is irrelevant in the RG sense, it suppresses thermal 
phase fluctuations and moves the phase boundary from $J_r^c$ 
at $T=0$ to smaller values of $J_r$ at finite temperatures.
Thus, for $\sigma<\sigma_c$ and $J_r$ less
than but close to the QCP, the system exhibits an interesting reentrant 
behavior (incoherent - coherent - incoherent) as the temperature is lowered.

The in-plane and the interlayer tunneling currents
behave qualitatively differently in the interlayer coherent phase. 
The pseudospin current is given by the inter-droplet current
$J_r\langle\sin(\theta_i-\theta_j)\rangle$ times the average number
of droplets normal to the direction of the
current flow. Separating $\theta$ into a fluctuating part $\tilde\theta$ 
and a slowly varying classical part $\theta_0$,
and average over $\tilde\theta$ within the above
meanfield theory, we obtain
$I_{2d}=2\sqrt{N_d}J_r\Psi^2(d\theta_0/dx)$ in the continuum limit,
where $N_d$ is the total number of droplets. Thus the in-plane critical
current increases monotonically with $\Psi$ as the temperature is lowered.
The superfluid stiffness can be identified as $\rho_s=J_r\Psi^2$.
The theory predicts that the tunneling current, on the other
hand, exhibits a non-monotonic temperature dependence.
In an ideal setup for tunneling, the current flows in the 
interlayer direction and is given by
$I_T=N_dt_r\langle\sin\theta\rangle=I_c\sin\theta_0$, where the
$I_c=N_d t(T\tau_0)^\mu\Psi$. The tunneling
current thus decreases initially with decreasing temperatures near
the transition. However, at low temperatures or well inside the
SC phase, the in-plane superfluid stiffness $\rho_s$ will cut off
the logarithmic singularity in the dissipation induced Debye-Waller factor 
for tunneling, $W_t$ in Eq.(\ref{wtlog}), 
such that $T$ is replace by $\rho_s$ in Eq.~(\ref{trpower}).
This leads to $I_c=N_d t(J_r\Psi^2\tau_0)^\mu\Psi$ which then grows
with the superfluid order parameter upon further lowering of the temperature.

In conclusion, the interplay between interlayer tunneling and 
quantum dissipation is studied in quantum Hall bilayers at $\nu=1$. 
We have shown the existence of a
critical dissipation separating a strong dissipation phase I 
from a weak dissipation phase II.
Phase I is an integer quantum Hall state driven by interlayer
tunneling and the development of a single-particle gap.
In phase II, the tunneling is renormalized to zero at low temperatures
by quantum fluctuations. 
This enables a quantum phase transition between
a compressible WC phase II(a) and a SC collective quantum Hall state II(b)
supported by spontaneous interlayer phase coherence.
The observation of the linearly dispersing Goldstone mode with in-plane
magnetic field \cite{spielman} is consistent with the
superfluid phase II(b). However, the fact that a weaker conductance peak
remains at zero-bias suggests that in-plane superfluidity is perhaps not the
only driving force behind interlayer coherence and the system might not be
far from phase I where interlayer tunneling itself is able to
produce coherent zero-bias tunneling. Future experiments on the
temperature dependence of the residual conductance peak may help 
resolve its origin. We have not addressed the
finite height and the nonzero width of the zero-bias conductance peak.
The lack of an apparent DC Josephson effect may be accounted for
by the absence of a finite temperature Kosterlitz-Thouless transition
due to nonzero interlayer tunneling and/or the presence of this and other 
forms of dissipation, such as the vortices induced by density inhomogeneities 
\cite{balents,stern,fogler,fertig2,joglekar} not considered in the 
present theory.

The author thanks L. Balents, J. Eisenstein, S.~M. Girvin,
and D.~N Sheng for discussions.
This work is supported in part by DOE grant No. DE-FG02-99ER45747 
and ACS grant No. 39498-AC5M.

\end{document}